\documentstyle[12pt]{article}
\begin{document}


\begin{center}
{\Large{\bf  Quantization of Space-time Noncommutative 
Theory\footnote{Talk given at SI2004, Fuji-Yoshida, Japan,
August 12-19, 2004 (to be published in the Proceedings).}}}
\end{center}
\vskip .5 truecm
\centerline{\bf Kazuo Fujikawa }
\vskip .4 truecm
\centerline {\it Institute of Quantum Science, College of 
Science and Technology  }
\centerline {\it Nihon University, Chiyoda-ku, Tokyo 101-8308, 
Japan}
\vskip 0.5 truecm

\makeatletter
\@addtoreset{equation}{section}
\def\theequation{\thesection.\arabic{equation}}
\makeatother

\begin{abstract}
General aspects of the quantization of field theories non-local 
in time are discussed. The path integral on the basis of 
Schwinger's action principle and the Bjorken-Johnson-Low  
prescription, which helps to recover the canonical structure 
from the results of the path integral, are used as the main 
machinery. A modified time ordering operation which formally
restores unitarity in field theories non-local 
in time is analyzed in detail. It is shown that the 
perturbative unitarity and the positive energy condition, in 
the sense that only the positive energy flows in the positive 
time direction for any fixed time-slice in space-time, are not 
simultaneously satisfied for theories non-local in time such as
space-time noncommutative theory. 
\end{abstract}

\section{Introduction}
 The quantization of  field theories non-local in time is 
problematic. First of all, no canonical formulation of such  
theories is known since a sensible definition of canonical 
momenta is not known, and the violation of unitarity appears to 
be the general aspects of such theories\cite{marnelius}. This 
issue has re-appeared in the recent study of space-time 
noncommutative theories\cite{gomis-mehen, alvarez-gaume, chu}. 
An interesting new development in this subject is that a 
suitable modified definition of time-ordering operation restores
 the perturbative unitarity in space-time noncommutative 
theory~\cite{bahns, rim-yee, sibold}, though its consistency 
with other basic postulates in conventional quantized field 
theory remains to be clarified.

We discuss the issues related to unitarity 
and the modified time ordering prescription on the basis of a 
recent paper\cite{fujikawa}, which analyzed these issues by 
using  path integral formulation and the Bjorken-Johnson-Low 
(BJL)\cite{bjorken} prescription.

\section{Quantization of higher derivative theory}

We present  a path integral formulation of 
higher derivative theory and then show how to recover the 
canonical structure from the path integral. 
For simplicity, we study the theory defined by
\begin{equation}
{\cal L}=\frac{1}{2}\partial_{\mu}\phi(x)\partial^{\mu}\phi(x)
+\lambda\frac{1}{2}\phi(x)\Box^{2}\phi(x)
\end{equation}
where\footnote{Our metric convention is 
$g_{\mu\nu}=(1,-1,-1,-1)$.}
$\Box=\partial_{\mu}\partial^{\mu}$
and $\lambda$ is a real constant. The canonical formulation
of higher derivative theory such as the present one has been 
analyzed in \cite{nambu}, for example.

We instead start with Schwinger's action principle and 
consider the Lagrangian with a source function $J(x)$
\begin{equation}
{\cal L}_{J}=\frac{1}{2}\partial_{\mu}\phi(x)
\partial^{\mu}\phi(x)
+\lambda\frac{1}{2}\phi(x)\Box^{2}\phi(x) +J(x)\phi(x).
\end{equation}
The Schwinger's action principle starts with the equation
of motion
\begin{eqnarray}
&&\langle +\infty|-\Box\hat{\phi}(x)+\lambda\Box^{2}
\hat{\phi}(x)+J(x)
|-\infty\rangle_{J}\nonumber\\
&&=\{-\Box\frac{\delta}{i\delta J(x)}
+\lambda\Box^{2}\frac{\delta}{i\delta J(x)}+J(x)\}
\langle +\infty|-\infty\rangle_{J}=0.
\end{eqnarray}
We here assume the existence of a formally quantized field
$\hat{\phi}(x)$, though its detailed properties are not specified
yet, and the asymptotic states $|\pm\infty\rangle_{J}$
in the presence of a source function $J(x)$ localized in 
space-time. The path
integral is then defined as a formal  solution of
the above functional equation
\begin{equation}
\langle +\infty|-\infty\rangle_{J}=
\int{\cal D}\phi\exp\{i\int d^{4}x{\cal L}_{J}\}.
\end{equation}
We now define the Green's function by
\begin{eqnarray}
\langle +\infty|T^{\star}\hat{\phi}(x)\hat{\phi}(y)
|-\infty\rangle
&=&\frac{\delta}{i\delta J(x)}\frac{\delta}{i\delta J(y)}
\langle +\infty|-\infty\rangle_{J}|_{J=0}\nonumber\\
&=&\frac{1}{i}\frac{1}{\Box - i\epsilon - 
\lambda\Box^{2}}\delta(x-y).
\end{eqnarray}
This Green's function contains all the information about the 
quantized field.

The BJL prescription states that we can replace the covariant 
$T^{\star}$ product by the conventional $T$ product when 
\begin{equation}
\lim_{k^{0}\rightarrow i\infty}\int d^{4}x e^{ik(x-y)}
\langle +\infty|T^{\star}\hat{\phi}(x)\hat{\phi}(y)
|-\infty\rangle=\lim_{k^{0}\rightarrow i\infty}
\frac{i}{k^{2} + i\epsilon + \lambda(k^{2})^{2}}=0.
\end{equation}
Thus we have
\begin{equation}
\int d^{4}x e^{ik(x-y)}
\langle +\infty|T \hat{\phi}(x)\hat{\phi}(y)
|-\infty\rangle=\frac{i}{k^{2}+ i\epsilon + \lambda(k^{2})^{2}}.
\end{equation}
By multiplying a suitable powers of the momentum variable 
$k_{\mu}$, we can recover the canonical commutation relations.
For example, 
\begin{eqnarray}
&&k_{\mu}\int d^{4}x e^{ik(x-y)}
\langle +\infty|T \hat{\phi}(x)\hat{\phi}(y)
|-\infty\rangle\nonumber\\
&&=\int d^{4}x( -i\partial^{x}_{\mu}e^{ik(x-y)})
\langle +\infty|T \hat{\phi}(x)\hat{\phi}(y)
|-\infty\rangle\nonumber\\
&&=\int d^{4}x e^{ik(x-y)}\{
\langle +\infty|T i\partial^{x}_{\mu} \hat{\phi}(x)\hat{\phi}(y)
|-\infty\rangle\nonumber\\ 
&&\ \ \ \ \ \ \ \ + i\delta(x^{0}-y^{0})
\langle +\infty|[\hat{\phi}(x),\hat{\phi}(y)]|-\infty\rangle
\}\nonumber\\
&&=\frac{ik_{\mu}}{k^{2}+ i\epsilon + \lambda(k^{2})^{2}}.
\end{eqnarray}
An analysis of this relation in the limit 
$k_{0}\rightarrow i\infty$  gives
\begin{eqnarray}
&&\delta(x^{0}-y^{0})[\hat{\phi}(x),\hat{\phi}(y)]=0,\nonumber\\
&&\int d^{4}x e^{ik(x-y)}\{
\langle +\infty|T i\partial^{x}_{\mu} \hat{\phi}(x)\hat{\phi}(y)
|-\infty\rangle 
=\frac{ik_{\mu}}{k^{2}+ i\epsilon + \lambda(k^{2})^{2}}.
\end{eqnarray}
Note that the limit  $k_{0}\rightarrow i\infty$ of the Fourier
transform of a $T$ product such as 
$\langle +\infty|T i\partial^{x}_{\mu} \hat{\phi}(x)\hat{\phi}(y)
|-\infty\rangle$ vanishes {\em by definition}.

By repeating the procedure with (2.9), we obtain
\begin{eqnarray}
&&\delta(x^{0}-y^{0})[\hat{\phi}(x),\hat{\phi}(y)]=0,\nonumber\\
&&\delta(x^{0}-y^{0})[\partial_{0}\hat{\phi}(x),\hat{\phi}(y)]
=0,\nonumber\\
&&\delta(x^{0}-y^{0})
[\partial^{2}_{0}\hat{\phi}(x),\hat{\phi}(y)]
=0,\nonumber\\
&&\delta(x^{0}-y^{0})
[\partial^{3}_{0}\hat{\phi}(x),\hat{\phi}(y)]
=\frac{i}{\lambda}\delta^{(4)}(x-y),
\nonumber\\
&&\delta(x^{0}-y^{0})[\partial_{0}\hat{\phi}(x),
\partial_{0}\hat{\phi}(y)]
=0,\nonumber\\
&&\delta(x^{0}-y^{0})
[\partial^{2}_{0}\hat{\phi}(x),\partial_{0}\hat{\phi}(y)]
=-\frac{i}{\lambda}\delta^{(4)}(x-y).
\end{eqnarray}
The general rule is that the commutator
\begin{equation}
[\hat{\phi}^{(m)}(x), \hat{\phi}^{(l)}(y)]\delta(x^{0}-y^{0})
\neq 0
\end{equation}
where $m+l=n-1$ for a theory with the n-th time derivative. 
Here  $\hat{\phi}^{(l)}(x)$ stands for the l-th time derivative
of $\hat{\phi}(x)$, $\hat{\phi}^{(l)}(x)
=\frac{\partial^{l}}{\partial(x^{0})^{l}}\hat{\phi}(x)$.
 We thus derive all the canonical commutation relations (2.10)
from the path integral defined by the  Schwinger's action 
principle and the $T^{\star}$ product, and those 
 commutation relations naturally agree with the relations 
derived by a canonical formulation of the higher derivative 
theory\cite{nambu}.

\section{Quantization of a  theory non-local in time}

We examine a non-local theory defined by 
\begin{eqnarray}
{\cal L}_{J}&=&-\frac{1}{2}\phi(x)\Box
[\phi(x+\xi)+\phi(x-\xi)] + J(x)\phi(x).
\end{eqnarray}
A formal integration of the Schwinger's action principle
gives a path integral
\begin{eqnarray}
\langle +\infty|-\infty\rangle_{J}=
\int{\cal D}\phi\exp\{i\int d^{4}x{\cal L}_{J}\},
\end{eqnarray}
which in turn leads to the correlation function
\begin{equation}
\int d^{4}x e^{ik(x-y)}
\langle T^{\star}\hat{\phi}(x)\hat{\phi}(y)\rangle =
\frac{i}{(k^{2}+ i\epsilon)[e^{ik\xi}+e^{-ik\xi}]}.
\end{equation}
For a time-like vector $\xi$, which may be chosen as 
$(\xi^{0},0,0,0)$,  the right-hand side of this
expression multiplied by any power of $k_{0}$ goes to zero 
for $k_{0}$ along the imaginary axis 
in the complex $k_{0}$ plane. Thus the application of BJL
prescription leads to (for any pair of non-negative integers $n$ 
and $m$) 
\begin{equation}
[\hat{\phi}^{(n)}(x), \hat{\phi}^{(m)}(y)]\delta(x^{0}-y^{0})=0
\end{equation}
where $\hat{\phi}^{(n)}(x)$ stands for the n-th time derivative
of $\hat{\phi}(x)$ as in (2.11).
 This relation is consistent with the $N\rightarrow\infty$ 
limit of a  higher derivative theory obtained by a truncation 
of the power series exapansion of $e^{\pm i\xi\hat{p}}$ at the 
$N$-th power in the starting Lagrangian (3.1).  

We next analyze a theory which contains a non-local interaction 
\begin{eqnarray}
{\cal L}&=&\frac{1}{2}\partial_{\mu}\phi(x)\partial^{\mu}\phi(x)
-\frac{m^{2}}{2}\phi(x)\phi(x)\nonumber\\
&& 
-\frac{g}{2}[\phi(x+\xi)\phi(x)\phi(x-\xi)+
\phi(x-\xi)\phi(x)\phi(x+\xi)]
+\phi(x) J(x)
\end{eqnarray}
where $\xi^{\mu}$ is a constant four-vector. This theory is 
not Lorentz invariant because of the constant vector $\xi^{\mu}$.
When one chooses $\xi^{\mu}$ to be 
$\xi^{\mu}=(\xi^{0},0,0,0)\ \  {\rm with}\ \ \xi^{0}>0$,
the quantization of the above theory  is analogous to that of 
 space-time noncommutative theory.

One may study the path integral quantization without specifying
  the precise quantization condition of field variables. 
One may thus  define a path integral by means of  Schwinger's 
action  principle and a suitable ansatz of asymptotic conditions
as in (3.2)
\begin{eqnarray}      
\langle +\infty|-\infty\rangle_{J}=\int{\cal D}\phi
\exp[i\int d^{4}x{\cal L}_{J}].
\end{eqnarray}
One may then define a formal expansion in powers of the coupling
constant $g$. We study one-loop diagrams in a formal 
perturbative expansion in powers of the coupling constant $g$ by
 starting with a tentative ansatz of quantization
\begin{equation}
\langle T^{\star}\hat{\phi}(x)\hat{\phi}(y)\rangle=
\frac{-i}{\Box + m^{2}-i\epsilon}\delta(x-y)
=\int\frac{d^{4}k}{(2\pi)^{4}}
e^{-ik(x-y)}\frac{i}{k^{2}-m^{2}+i\epsilon} 
\end{equation}
which is equivalent to a canonical quantization of free theory.
One-loop self-energy diagrams contain the contribution
\begin{eqnarray}
\frac{(-ig)^{2}}{2}\int d^{4}x d^{4}y \phi(x)\phi(y)
\langle T^{\star}\phi(x+\xi)\phi(y-\xi)\rangle\langle
T^{\star}\phi(x-\xi)\phi(y-2\xi)\rangle
\end{eqnarray}
which contains the finite non-local terms separated  
up to the order of $\sim 3\xi$.

This  term in (3.8) gives rise to  
\begin{eqnarray}
&&g^{2}i\Sigma(k, \xi)\nonumber\\
&=&\frac{ig^{2}}{2(4\pi)^{2}}
\int_{0}^{1}dx[e^{i(2-x)\xi p} + e^{-i(2-x)\xi p}]
\int_{0}^{\infty}
\frac{d\alpha}{\alpha}e^{i\alpha[x(1-x)p^{2}-m^{2}+i\epsilon)]
-i\frac{\xi^{2}}{4\alpha}}.
\end{eqnarray}
Following the conventional approach, we define the integral
for a Euclidean momentum $p_{\mu}$, for which $p^{2}<0$.
In this case, one can deform the integration contour in the 
variable $\alpha$ along the negative real axis and obtain
\begin{eqnarray}
\int_{0}^{\infty}
\frac{d\alpha}{\alpha}e^{i\alpha[x(1-x)p^{2}-m^{2}+i\epsilon)]
-i\frac{\xi^{2}}{4\alpha}}=
-i\pi H^{(2)}_{0}(-i
\sqrt{(-\xi^{2})[-x(1-x)p^{2}+m^{2}-i\epsilon)})
\end{eqnarray}
for $\xi^{2}<0$. Here $H^{(2)}_{0}(z)$ stands for the Hankel
function which has an asymptotic expansion for $|z|\rightarrow
\infty$
\begin{eqnarray}
H^{(2)}_{0}(z)\sim \sqrt{\frac{2}{\pi z}}e^{-i(z-\frac{\pi}{4})}
\end{eqnarray}
for $-2\pi <{\rm arg} z <\pi$.

We thus find that for $p_{0}\rightarrow i\infty$ 
\begin{eqnarray}
\int_{0}^{\infty}
\frac{d\alpha}{\alpha}e^{i\alpha[x(1-x)p^{2}-m^{2}+i\epsilon)]
-i\frac{\xi^{2}}{4\alpha}}\sim
-\pi \sqrt{\frac{2}{\pi z}}e^{-z}
\end{eqnarray}
with $z=\sqrt{(-\xi^{2})[-x(1-x)p^{2}+m^{2}-i\epsilon)}$
for a space-like $\xi$, $\xi^{2}<0$. On the other hand, we have 
a damping oscillatory behavior for 
$p_{0}\rightarrow i\infty$, 
\begin{eqnarray}
\int_{0}^{\infty}
\frac{d\alpha}{\alpha}e^{i\alpha[x(1-x)p^{2}-m^{2}+i\epsilon)]
-i\frac{\xi^{2}}{4\alpha}}\sim
-i\pi \sqrt{\frac{2}{\pi z}}e^{(- iz+i\frac{\pi}{4})}
\end{eqnarray}
with $z=\sqrt{(\xi^{2})[-x(1-x)p^{2}+m^{2}-i\epsilon)}$
for a time-like $\xi$, $\xi^{2}>0$, which is defined by an 
analytic continuation.

When one writes the (complete) connected two-point correlation 
function with one-loop corrections as 
\begin{equation}
\langle T^{\star}\hat{\phi}(x)\hat{\phi}(y)\rangle_{ren}
=\int\frac{d^{4}p}{(2\pi)^{4}}
e^{-ip(x-y)}\frac{i}{p^{2}+g^{2}\Sigma(p, \xi)-m_{r}^{2}
+i\epsilon}, 
\end{equation}
the two-point function generally contains the
 non-local term in $g^{2}\Sigma(p, \xi)$. 
When one applies the BJL prescription to the two-point 
correlation function, for the time-like $\xi$ for which we may 
take $\xi=(\xi^{0}, \vec{0})$, $g^{2}\Sigma(p, \xi)$ diverges 
exponentially for $p_{0}\rightarrow i\infty$. This arises from
the behavior of the factor   
$[e^{i(2-x)\xi p} + e^{-i(2-x)\xi p}]$
in (3.9) for $p_{0}\rightarrow i\infty$ and 
$\xi=(\xi^{0}, \vec{0})$, which dominates the damping 
oscillatory behavior (3.13)\footnote{The non-vanishing imaginary
part of $g^{2}\Sigma(p, \xi)$ in (3.13) for the Euclidean 
momentum given by $p_{0}\rightarrow i\infty$ is associated with 
the violation of unitarity in the present theory non-local in 
time~\cite{gomis-mehen, alvarez-gaume, chu}.}. The canonical 
structure specified 
by the BJL analysis is thus completely modified by the one-loop 
effects of  the  interaction non-local in time. After one-loop 
corrections, we essentially have the same result (3.3) as for 
the non-local theory (3.1). 
The naive ansatz of the two-point correlation function at the 
starting point of  perturbation theory (3.7) is not justified. 

Nevertheless, it is instructive to examine the formal 
perturbative unitarity of an S-matrix defined for the theory 
non-local in time.  One may start with a naive Hamiltonian
\begin{eqnarray}
{\cal H}&=&\frac{1}{2}\Pi^{2}(x)
+\frac{1}{2}\vec{\nabla}\phi(x)\vec{\nabla}\phi(x)    
+\frac{1}{2}m^{2}\phi^{2}(x)\nonumber\\
&& +\frac{g}{2}[\phi(x-\xi)\phi(x)\phi(x+\xi)+
\phi(x+\xi)\phi(x)\phi(x-\xi)]
\end{eqnarray}
where $x^{\mu}=(0,\vec{x})$ and $\xi^{\mu}=(\xi^{0},\vec{0})$
, and $\Pi(x)=\frac{\partial}{\partial x^{0}}\phi(x)$ is a naive
 canonical momentum conjugate to $\phi(x)$.
This Hamiltonian is formally hermitian, 
${\cal H}^{\dagger}={\cal H}$, but ${\cal H}$ is not local in 
the time coordinate and does not generate time development in 
the conventional sense for the small time interval 
$\Delta t<\xi^{0}$.

One may then observe that 
\begin{eqnarray}
S(t_{+}, t_{-})=e^{i\hat{H}_{0}t_{+}}e^{-i\hat{H}(t_{+}-t_{-})}
e^{-i\hat{H}_{0}t_{-}}
\end{eqnarray}
for ${\cal H}$ in (3.6) with $H_{0}$ the free Hamiltonian
is unitary, $S(t_{+}, t_{-})^{\dagger}S(t_{+}, t_{-})
=S(t_{+}, t_{-})S(t_{+}, t_{-})^{\dagger}\\ =1$.
The formal power series expansion in the coupling 
constant\footnote{We use the notation $T_{\star}$ for the time 
ordering in non-local theory, whereas $T$ or $T^{\star}$ is 
used for the conventional time ordering with respect to 
the time variable of {\em each} field variable $\phi(x)$. }
\begin{eqnarray}
S(t_{+}, t_{-})=\sum_{n=0}^{\infty}\frac{(-i)^{n}}{n!}
\int_{t_{-}}^{t_{+}}dt_{1}....\int_{t_{-}}^{t_{+}}dt_{n}
T_{\star}(\hat{H}_{I}(t_{1})....\hat{H}_{I}(t_{n}))
\end{eqnarray}
with a hermitian 
\begin{equation}
\hat{H}_{I}(t)=e^{i\hat{H}_{0}t}\int d^{3}x
\frac{g}{2}[\phi(-\xi,\vec{x})\phi(0,\vec{x})\phi(\xi,\vec{x})
+\phi(\xi,\vec{x})\phi(0,\vec{x})\phi(-\xi,\vec{x})]
e^{-i\hat{H}_{0}t}
\end{equation}
thus defines a unitary S-matrix for $t_{-}\rightarrow -\infty, 
t_{+}\rightarrow +\infty$.
This definition of a unitary operator corresponds to the 
definition of a unitary S-matrix for space-time noncommutative 
theory proposed in~\cite{bahns, rim-yee}. 

It is important to recognize that the time-ordering in the 
present context is defined with repect to the time variable
appearing in $\hat{H}_{I}(t)$; if one performs a time-ordering
with repect to the time variable appearing in each field variable
$\phi(x)$, one generally obtains different results due to the 
non-local structure of the interaction term in time. Since the 
operator $\hat{S}$ defined above is manifestly unitary, the 
non-unitary result in the conventional Feynman rules, which are 
based on the time-ordering of each operator $\phi(x)$, arises 
from this difference of time ordering. 

When one defines 
\begin{eqnarray}
A_{1}&=&\int_{-\infty}^{\infty} dt \hat{H}_{I}(t),\nonumber\\ 
A_{2}&=&\frac{1}{2}\int_{-\infty}^{\infty} 
\int_{-\infty}^{\infty} 
dt_{1}dt_{2}T_{\star}\hat{H}_{I}(t_{1})\hat{H}_{I}(t_{2})
\nonumber\\
&=&\int_{-\infty}^{\infty} 
\int_{-\infty}^{\infty} 
dt_{1}dt_{2}\theta(t_{1}-t_{2})
\hat{H}_{I}(t_{1})\hat{H}_{I}(t_{2})
\end{eqnarray}
the unitarity relation of the above S-matrix in the second order
 of the coupling constant requires( see, for example, 
\cite{rim-yee})
\begin{eqnarray}
A_{2}+A_{2}^{\dagger}=A_{1}^{\dagger}A_{1}=A_{1}^{2}.
\end{eqnarray}
To be explicit
\begin{eqnarray}
A_{2}+A_{2}^{\dagger}&=&\int_{-\infty}^{\infty} 
\int_{-\infty}^{\infty} 
dt_{1}dt_{2}\theta(t_{1}-t_{2})\{
\hat{H}_{I}(t_{1})\hat{H}_{I}(t_{2})
+\hat{H}_{I}(t_{2})\hat{H}_{I}(t_{1})\}\nonumber\\
&=&\int_{-\infty}^{\infty} 
\int_{-\infty}^{\infty} 
dt_{1}dt_{2}\{\theta(t_{1}-t_{2})
\hat{H}_{I}(t_{1})\hat{H}_{I}(t_{2})
+\theta(t_{2}-t_{1})\hat{H}_{I}(t_{1})\hat{H}_{I}(t_{2})\}
\nonumber\\
&=&\int_{-\infty}^{\infty}  dt_{1}\hat{H}_{I}(t_{1})
\int_{-\infty}^{\infty}dt_{2}\hat{H}_{I}(t_{2})=A_{1}^{2}
\end{eqnarray}
by noting $\theta(t_{1}-t_{2})+\theta(t_{2}-t_{1})=1$, as 
required by the unitarity relation.

In contrast, if one uses the conventional time ordering one has
\begin{eqnarray} 
A_{2}&=&\frac{1}{2}\int_{-\infty}^{\infty} 
\int_{-\infty}^{\infty} 
dt_{1}dt_{2}T^{\star}\hat{H}_{I}(t_{1})\hat{H}_{I}(t_{2})
\nonumber\\
&\neq&\int_{-\infty}^{\infty} 
\int_{-\infty}^{\infty} 
dt_{1}dt_{2}\theta(t_{1}-t_{2})
\hat{H}_{I}(t_{1})\hat{H}_{I}(t_{2})
\end{eqnarray}
since the time ordering by $T^{\star}$ is defined with respect 
to the time variable of each field $\phi(t,\vec{x})$, and thus 
the unitarity of the conventional operator 
\begin{eqnarray}
\hat{S}=\sum_{n=0}^{\infty}\frac{(-i)^{n}}{n!}
\int_{-\infty}^{+\infty}dt_{1}....\int_{-\infty}^{+\infty}
dt_{n}T^{\star}(\hat{H}_{I}(t_{1})....\hat{H}_{I}(t_{n}))
\end{eqnarray}
is not satisfied for the non-local $\hat{H}_{I}(t)$ in general.
Note that the perturbative expansion with the $T^{\star}$
product is directly defined by the path integral without 
recourse to the expression such as (3.15).

On the other hand, the positive energy condition,
which is ensured by the Feynman propagator, is  not obvious for 
the propagator defined by the modified time ordering $T_{\star}$.
 To be specific, we have the following correlation function in 
the Wick-type reduction of the S-matrix in (3.23)
\begin{eqnarray}
&&\langle0|T\phi(x-\xi)\phi(y+\xi)|0\rangle\nonumber\\
&&=
\int\frac{d^{4}k}{(2\pi)^{4}}\exp[-ik((x-\xi)-(y+\xi))]
\frac{i}{k_{\mu}k^{\mu}-m^{2}+i\epsilon}\\
&&=\theta((x-\xi)^{0}-(y+\xi)^{0})
\int\frac{d^{3}k}{(2\pi)^{3}2\omega}
\exp[-i\omega((x-\xi)^{0}-(y+\xi)^{0})+
i\vec{k}(\vec{x}-\vec{y})]\nonumber\\
&&+\theta((y+\xi)^{0}-(x-\xi)^{0})
\int\frac{d^{3}k}{(2\pi)^{3}2\omega}
\exp[-i\omega((y+\xi)^{0}-(x-\xi)^{0})+
i\vec{k}(\vec{y}-\vec{x})]
\nonumber
\end{eqnarray}
with $\omega=\sqrt{\vec{k}^{2}+m^{2}}$ for the conventional 
Feynman prescription with $m^{2}-i\epsilon$, which ensures that 
the positive frequency components propagate in the forward time 
direction and the negative frequency components propagate in 
the backward time direction and thus the positive energy  
flows always in the forward time 
direction. The path integral with respect to the 
field variable $\phi(x)$ gives this time ordering
 or the $T^{\star}$ product.

In comparison, the non-local prescription (3.17) gives the 
following correlation function for the quantized free field in 
the Wick-type reduction
\begin{eqnarray}
&&\langle0|T_{\star}\phi(x-\xi)\phi(y+\xi)|0\rangle\nonumber\\
&&=\theta(x^{0}-y^{0})
\int\frac{d^{3}k}{(2\pi)^{3}2\omega}
\exp[-i\omega((x-\xi)^{0}-(y+\xi)^{0}))+
i\vec{k}(\vec{x}-\vec{y})]\nonumber\\
&&+\theta(y^{0}-x^{0})
\int\frac{d^{3}k}{(2\pi)^{3}2\omega}
\exp[-i\omega((y+\xi)^{0}-(x-\xi)^{0})+
i\vec{k}(\vec{y}-\vec{x})]
\end{eqnarray}
where the time-ordering step function $\theta(x^{0}-y^{0})$,
for example, and the signature of the time variable 
$(x-\xi)^{0}-(y+\xi)^{0}$ appearing in the exponential are
not correlated, and it 
 allows the negative energy to propagate in the forward
time direction also. This result is not reproduced by the 
Feynman's $m^{2} - i\epsilon$ prescription. When one considers 
an arbitrary fixed time-slice in 4-dimesional space-time, the 
condition that
all the particles crossing the time-slice carry the positive 
energy in the forward
time direction, which is regarded as the positive energy 
condition in the path integral formulation\cite{spin-sta} 
(or in perturbation theory in general), is not satisfied. This 
positive energy condition  is crucial in the analysis of 
spin-statistics theorem\cite{spin-sta, pauli}, for example.

We thus summarize the analysis of  this section as follows:
The naive canonical quantization in a perturbative sense is not 
justified in the present theory non-local in time
when one incorporates the higher order corrections. The unitarity
of the (formal) perturbative S-matrix is ensured if one adopts 
the $T_{\star}$ product, but the positive energy condition is not
satisfied by this prescription. On the other hand,
the unitarity of the $S$ matrix is spoiled if one adopts the 
conventional $T$ or $T^{\star}$ product which is defined by the 
path integral, though the positive energy condition and a 
smooth Wick rotation are ensured.

\section{Space-time noncommutative theory}
We study the simplest noncommutative theory defined by
\begin{eqnarray}
{\cal L}_{J}
&=&\frac{1}{2}\partial_{\mu}\phi(x)\star\partial^{\mu}\phi(x)
-\frac{m^{2}}{2}\phi(x)\star\phi(x)\nonumber\\
&& 
-\frac{g}{3!}\phi(x)\star\phi(x)\star\phi(x)
+\phi(x)\star J(x)
\end{eqnarray}
where the $\star$ product is defined by the so-called Moyal
product
\begin{equation}
\phi(x)\star\phi(x)
=e^{\frac{i}{2}\xi\partial^{x}_{\mu}
\theta^{\mu\nu}\partial^{y}_{\nu}}
\phi(x)\phi(y)|_{y=x}
=e^{\frac{i}{2}\xi\partial_{x}\wedge\partial_{y}}
\phi(x)\phi(y)|_{y=x}.
\end{equation}
The real positive parameter $\xi$ stands for the deformation 
parameter,
and the antisymmetric parameter 
$\theta^{\mu\nu}=-\theta^{\nu\mu}$ corresponds to  
$i\xi\theta^{\mu\nu}=[\hat{x}^{\mu},\hat{x}^{\nu}]$ ; since this
 theory is  not Lorentz covariant we consider the case 
$\theta^{0i}=-\theta^{i0}\neq 0$ for a suitable $i$
but all others $\theta^{\mu\nu}=0$ in the following.
The path integral is then defined as a formal integral of the 
Schwinger's action principle
\begin{eqnarray}      
\langle +\infty|-\infty\rangle_{J}=\int{\cal D}\phi
\exp[i\int d^{4}x{\cal L}_{J}].
\end{eqnarray}

It has been argued \cite{chepelev} that the present theory is 
renormalizable in the formal perturbative expansion in powers of
 the coupling constant $g$ starting with 
\begin{equation}
\langle T^{\star}\hat{\phi}(x)\hat{\phi}(y)\rangle=
\frac{-i}{\Box + m^{2}-i\epsilon}\delta(x-y)
=\int\frac{d^{4}k}{(2\pi)^{4}}
e^{-ik(x-y)}\frac{i}{k^{2}-m^{2}+i\epsilon}
\end{equation}
which is equivalent to a canonical quantization of free theory.
The one-loop self-energy is given by
\begin{eqnarray}
&&g^{2}i\Sigma(p, \xi)\nonumber\\
&&=\frac{-g^{2}}{2}\int \frac{d^{4}k}{(2\pi)^{4}} 
\cos^{2}(\frac{\xi}{2}p\wedge k)\frac{i}{
((p-k)^{2}-m^{2}+i\epsilon)}\frac{i}{(k^{2}-m^{2}+i\epsilon)}
\nonumber\\
&&=\frac{g^{2}}{4}\int \frac{d^{4}k}{(2\pi)^{4}} 
\frac{1+\cos(\xi p\wedge k) }{((p-k)^{2}-m^{2}+i\epsilon)
(k^{2}-m^{2}+i\epsilon)}
\end{eqnarray}
Since the term without the factor $\cos(\xi p\wedge k)$ is 
identical to the conventional theory, we concentrate on the term
 with $\cos(\xi p\wedge k)$
\begin{eqnarray}
&&\frac{-g^{2}}{8}\int \frac{d^{4}k}{(2\pi)^{4}}
[e^{i\xi p\wedge k}+ e^{-i\xi p\wedge k}]
\int_{0}^{\infty}\alpha d\alpha\int_{0}^{1}dx
e^{i\alpha[k^{2}+x(1-x)p^{2}-m^{2}+i\epsilon]}
\nonumber\\
&&=\frac{i g^{2}}{4(4\pi)^{2}}
\int_{0}^{\infty}\frac{d\alpha}{\alpha}\int_{0}^{1}dx
e^{i\alpha[x(1-x)p^{2}-m^{2}+i\epsilon]-i\frac{\xi^{2}
\tilde{p}^{2}}{4\alpha}}
\end{eqnarray}
where
\begin{equation}
\tilde{p}^{\mu}=\theta^{\mu\nu}p_{\nu}.
\end{equation}
See also (3.10).
For space-time noncommutative theory with $\theta^{01}\neq 0$, 
for example, $\tilde{p}^{2}\sim p_{1}^{2}-p_{0}^{2}$,
and for space-space noncommutative theory with 
$\theta^{23}\neq 0$, for example, $\tilde{p}^{2}\sim 
-p_{2}^{2}-p_{3}^{2}$.
We thus obtain by using the result in (3.10)
\begin{eqnarray}
&&g^{2}i\Sigma(p, \xi)_{non-planar}\nonumber\\
&&=\frac{\pi g^{2}}{4(4\pi)^{2}} H^{(2)}_{0}(-i
\sqrt{(-\xi^{2}\tilde{p}^{2})[-x(1-x)p^{2}+m^{2}-i\epsilon)})
\end{eqnarray}
for $\xi^{2}\tilde{p}^{2}<0$, namely, for space-space 
noncommutative theory. For space-time noncommutative theory, for
which $\xi^{2}\tilde{p}^{2}$ can be positive as well as 
negative, one defines the amplitude by an analytic continuation. 

As for the consistency of the naive quantization (4.4), it is 
important to analyze the self-energy correction in 
\begin{equation}
\langle T^{\star}\hat{\phi}(x)\hat{\phi}(y)\rangle_{ren}
=\int\frac{d^{4}p}{(2\pi)^{4}}
e^{-ip(x-y)}\frac{i}{p^{2}+g^{2}\Sigma(p, \xi)-m_{r}^{2}}. 
\end{equation}
If $\Sigma(p, \xi)$ contains a non-local exponential factor,
the naive quantization is not justified in the framework of the 
BJL prescription as we explained for 
the simple theory non-local in time in the previous section.
By using the asymptotic expansion in (3.11), 
it is shown that the naive quantization (4.4), either in 
space-space or in space-time noncommutative theory, is not 
modified by the one-loop corrections in the framework of BJL
prescription. This is in sharp contrast to the simple theory
non-local in time analyzed in the previous section. This 
difference arises from the fact that $p_{\mu}\theta^{\mu\nu}
p_{\nu}=0$
and thus the two-point function, which depends on the single
momentum $p_{\mu}$, does not contain an extra
exponential factor in the present space-time noncommutative
theory. Although our analysis does not justify the naive 
quantization to the non-perturbative accuray, it provides a 
basis of the formal perturbative expansion in the present 
model\cite{filk}.

As for the analysis of unitarity, the argument is almost 
identical to that for the simple non-local theory in the previous
section. The one-loop self-energy amplitude in the path 
integral formulation (4.6) is obtained from (3.9) by replacing 
\begin{eqnarray}
\xi^{2} \rightarrow \xi^{2}\tilde{p}^{2}
\end{eqnarray}
except for the exponential prefactor. The amplitude for the
Euclidean momentum exhibits  oscillatory behavior with a 
non-vanishing imaginary part for the space-time noncommutative
theory; this indicates the violation of 
unitarity~\cite{gomis-mehen, alvarez-gaume, chu} in the path
integtal formulation with the conventional time-ordering. On
the other hand, the amplitude for the space-space noncommutative
theory is real and exponentially damping for the Euclidean 
momentum, and thus no violation of unitarity in the conventional
 time-ordering. See also (3.12) and (3.13).

On the other hand, the analysis of unitarity on the basis of 
modified time-ordering~\cite{bahns,rim-yee} starts with  
\begin{eqnarray}
S(t_{+}, t_{-})=e^{i\hat{H}_{0}t_{+}}e^{-i\hat{H}(t_{+}-t_{-})}
e^{-i\hat{H}_{0}t_{-}}
\end{eqnarray}
which is unitary
\begin{equation}
S(t_{+}, t_{-})^{\dagger}S(t_{+}, t_{-})
=S(t_{+}, t_{-})S(t_{+}, t_{-})^{\dagger}=1
\end{equation}
where the total Hamiltonian $\hat{H}=\int d^{3}x{\cal H}$ is 
defined by 
\begin{eqnarray}
{\cal H}&=&\frac{1}{2}\Pi^{2}(0,\vec{x})
+\frac{1}{2}\vec{\nabla}\phi(0,\vec{x})\vec{\nabla}
\phi(0,\vec{x})+\frac{1}{2}m^{2}\phi^{2}(0,\vec{x})\nonumber\\
&+&\frac{g}{2\cdot 3!}[\phi(0,\vec{x})\star\phi(0,\vec{x})\star
\phi(0,\vec{x}) + h.c. ]
\end{eqnarray}
with the naive canonical momentum $\Pi(x) 
=\frac{\partial}{\partial x^{0}}\phi(x)$ conjugate to the 
variable $\phi(x)$. See \cite{gomis} for a different approach 
to the Hamiltonian formulation of space-time noncommutative 
theory. 

The modified time ordering for the non-local Hamiltonian (4.13),
which defines a unitary S-matrix (4.11) in a perturbative 
expansion, however spoils the positive energy 
condition~\cite{fujikawa} in the sense that only the positive 
energy flows in the positive time direction for any fixed 
time-slice in space-time, just as (3.25) in the naive theory 
non-local in time in the previous section.

\section{Conclusion}

We discussed some of the basic aspects of quantized theory
which is non-local in the time variable on the basis of path
integral quantization. We analyzed the recent proposal of the 
modified time ordering prescription\cite{bahns,rim-yee}, which 
generally defines a unitary S-matrix for theories non-local in 
time. It has been shown that the unitary S-matrix has 
certain advantages but at the time it has several disadvantages,
 and the perturbative positive energy condition and Wick 
rotation to Euclidean thoery, which are ensured by the 
Feynman's $m^{2}-i\epsilon$ prescription in the path integral, 
are spoiled. The modified time ordering needs to be examined 
further whether it is consistent with other basic postulates in 
quantized field theory.

\end{document}